\documentclass[aps,pre,reprint, amsmath, amssymb,superscriptaddress]{revtex4-1}

\usepackage{morefloats}
\usepackage{bm}
\newcommand{\beq}{\begin{equation}}
\newcommand{\eeq}{\end{equation}}

\usepackage[retainorgcmds]{IEEEtrantools}
\usepackage{graphicx,tikz,placeins}
\usepackage{mathrsfs}
\usepackage{amsmath,amssymb,amsfonts,physics}
\usepackage{color}
\usepackage{float}
\usepackage{times,txfonts}
\usepackage{nicefrac}
\usepackage[colorlinks=true,linkcolor=blue,urlcolor=black,citecolor=blue,pdfusetitle]{hyperref}
\usepackage{physics}
\usepackage{soul}
\definecolor{fe}{HTML}{812fab}

\begin{document}
\title{ Emergent collective heat engines  from neighborhood-dependent thermal reservoirs}
\author{Carlos E. Fiore}
\affiliation{Universidade de São Paulo, Instituto de Física, Rua do Matão, 1371, 05508-090 São Paulo, SP, Brazil}
\date{\today}

\begin{abstract} 

We introduce and analyze a class of heat engines composed of interacting units, in which the thermal reservoir is associated with the neighborhood surrounding each unit.
These systems can be mapped onto stochastic opinion models and are characterized by collective behavior at low temperatures, as well as different types of phase transitions, marked by spontaneous symmetry breaking and classifications that depend on topology and neighborhood.
For the case of contact with two thermal baths—equivalent to each unit having $k = 4$ nearest neighbors—the system can be tuned to operate at maximum power without sacrificing efficiency or increasing dissipation. These quantities are related by the general expression $
\beta_2 {\cal P} \eta_c =-J_4 \eta \sigma$ when the worksource stems from different interaction energies.
The heat engine placed in contact with more than three reservoirs is  more revealing, showing that the intermediate thermal reservoir can be conveniently adjusted to achieve the desired compromise between power, efficiency, and dissipation. The influence of lattice topology (regular and random-regular networks), its relationship with collective operation  as well as the distinct ratios between the temperatures of the thermal baths, has also been investigated.

\end{abstract}

\maketitle

\section{Introduction}
Stochastic thermodynamics \cite{PhysRevE.82.021120,
esposito2012stochastic, seifert2012stochastic, tome2015stochastic} provides a general and unified framework
for addressing central issues in thermodynamics \cite{prigo, groot, broeck15}. It not only extends the laws of thermodynamics to nonequilibrium systems operating at the nanoscale but also offers alternative interpretations, such as fluctuation theorems \cite{j1, j2, kau, barau} and thermodynamic uncertainty relations (TURs) \cite{barato2015thermodynamic}.
The entropy production is a central quantity in stochastic thermodynamics
\cite{schnakenberg}, distinguishing equilibrium 
($\sigma = 0$) from nonequilibrium ($\sigma > 0$) systems. More recently, it has also been used as an indicator of nonequilibrium phase transitions \cite{noa2019entropy, martynec2020entropy, fiore2021current, basile}.

We now turn to the thermodynamics of heat engines.
This topic has aroused considerable interest, both because it extends the fundamental idea of energy conversion to the nanoscopic scale and because, in contrast to equilibrium thermodynamics, fluctuations in quantities and currents can become important \cite{forao2025characterization}, revealing that the choice of approach or protocol is crucial for ensuring the desired performance.
Generally, heat engines can be grouped into different categories, such as those operating under fixed conditions \cite{gatien, powerful, mamede23}, periodically driven systems \cite{proesmans2016brownian, mamede2021obtaining}, or even a sequential/cyclic description, in which the system is placed in contact with a single thermal reservoir (isothermal process), and an instantaneous switching (adiabatic process) allows the change of thermal reservoir \cite{noa2020thermodynamics, noa2021efficient, harunari2020maximal}.

We introduce an alternative design of heat engines, based on the idea of collective behavior among units, in which a well-defined thermal reservoir is associated with each local neighborhood surrounding the units. Such a description can be mapped onto opinion systems \cite{tome2022stochastic, felipe1}, for which a consistent thermodynamic framework was recently proposed, satisfying fluctuation theorems \cite{crooks1999entropy}.
Opinion models are crucial in various nonequilibrium systems, such as complex social processes, population dynamics, decision-making, elections, the spreading of fake news/rumors, and others \cite{castellano2009statistical}. Their simplest forms are those composed of two states per agent—the voter \cite{liggett2013stochastic} and majority vote (MV) models \cite{de1992isotropic}—which constitute remarkable examples and exhibit universal features of nonequilibrium phase transitions.
The simplest case (the MV model) involves an interaction mechanism in which an agent tends to align (follow) its opinion based on the majority opinion of its nearest neighbors \cite{de1992isotropic, pereira2005majority, PhysRevE.91.022816}.
Subsequent generalizations of the MV model have attracted significant interest, taking into account the influence of network topology \cite{pereira2005majority, PhysRevE.91.022816}, the inclusion of distinct types of noise \cite{vieira2016phase, encinas2019majority}, more states per agent \cite{brunstein99, PhysRevE.95.042304}, inertial effects \cite{PhysRevE.95.042304, harunari2017partial, encinas2018fundamental}, and other cooperative effects \cite{oliveira2024entropy}.

We perform a careful analysis, taking into account the influence of distinct elements such as the worksource, neighborhood, interaction topology and the relationship between temperatures. Our results show that the present approach is capable of designing heat engines that operate at maximum power with desired efficiency and controlled dissipation.

 This paper is organized as follows: Model thermodynamics, the neighborhood
 thermal reservoirs  and its relationship with opinion models are presented
 in Sec.~\ref{sec2}. The main results for two and three thermal baths
 are shown in Sec.~\ref{main} and conclusions are addressed in Sec.~\ref{cond}.

\section{Model and Thermodynamics}\label{sec2}
The systems we are dealing with are defined on a lattice of size $N$, and  
each microscopic configuration $\eta$ corresponds to a collection of  
$N$ spins, $\eta \equiv (\eta_1, \eta_2, \ldots, \eta_i, \ldots, \eta_N)$, with  
$\eta_i$ denoting the spin variable at site $i$, which takes the values $\pm 1$  
depending on whether the spin is ``up'' ($\eta_i = 1$) or ``down'' ($\eta_i = -1$).  
Its dynamics is governed by the following master equation:
\begin{equation}
\frac{d}{dt}P(\eta,t) = \sum_{i=1}^N \{w_i(\eta^i)P(\eta^i,t)
- w_i(\eta)P(\eta,t)\},
\label{eq3}
\end{equation}
where $w_i(\eta)$ comprises the transition rate at which each site $i$ changes its spin from $\eta_i$ to $-\eta_i$.  
The thermodynamic description is set up by assuming that each transition rate $w_i(\eta)$ can be decomposed into  
$k/2$ distinct (and mutually exclusive) components, each associated with a given \emph{thermal reservoir}.  
From Eq.~(\ref{eq3}), $w_i(\eta)$ is then written as  
\[
w_i(\eta) = \sum_\ell w_{\ell i}(\eta) \quad (\ell=2,4,\ldots,k),
\]
where each term $w_{\ell i}(\eta)$ assumes the Glauber form:

\begin{equation}
    w_{\ell i}(\eta)=\frac{\alpha_\ell}{2}\{1-\tanh\left[\frac{\beta_\ell}{2} (\Delta E_\ell-\theta_\ell F_\ell)\right] \},
    \label{eqg}
\end{equation}
where $\alpha_\ell$ is a constant, $\Delta E_\ell = E_\ell(\eta^i) - E_\ell(\eta)$ denotes  
the energy difference between configurations $\eta$ and $\eta^i$, which are placed in contact with the $\ell$-th thermal reservoir, with reciprocal inverse temperature $\beta_\ell$,  
and $\theta_\ell F_\ell$ accounts for the contribution of a work source  
with strength $F_\ell$, expressed in terms of the variable $\theta_\ell = \pm 1$.  
Our reservoir approach  
implies that all neighborhoods with the same $|\ell|$ have the same temperature $\beta_\ell$. For systems with "up-down" $Z_2$ symmetry, the energy can be generically expressed according to the Ising-like  
form  \cite{yeomans1992statistical}  
\[
E_\ell(\eta) = -J_\ell \sum_{(i,j)} \eta_i \eta_j - H_\ell \sum_{i=1}^N \eta_i,
\]
where $J_\ell$ represents the interaction energy over a neighborhood of $k$ spins and  
$H_\ell$ is the magnetic field.  
The system presents two ferromagnetic phases at low temperatures, and a  
ferromagnetic-paramagnetic phase transition takes place as the temperature is raised.  

We now proceed with the thermodynamic description. Starting  with the entropy definition $S=-\langle\ln P(\eta)\rangle$ (we adopt the convention 
$k_{B}=1$), its time derivative $dS/dt$ leads to the following  expression for the entropy production  in the nonequilibrium steady state (NESS), $P(\eta,t) \rightarrow p^{\rm st}(\eta)$ \cite{noa2019entropy}:
\begin{equation}
  \sigma = \sum_{i} \left\langle w_i(\eta) \ln \left( \frac{w_i(\eta)}{w_i(\eta^{i})} \right) \right\rangle
  \label{eq:ep3}
\end{equation}


The first law of Thermodynamics is established by taking the time derivative of the mean energy $U = \langle E(\eta) \rangle$,  
given by $dU/dt = \sum_\ell \Phi_\ell + {\cal P}$, where $\Phi_\ell$ denotes the heat exchanged  
with the $\ell$-th thermal reservoir, defined as
\begin{eqnarray}
\label{4}
\Phi_\ell &=&
\sum_i \langle [E_\ell(\eta^i) - E_\ell(\eta) - \theta_\ell F_\ell] w_{\ell i}(\eta) \rangle.
\end{eqnarray}
The expression for the power ${\cal P}$ can be obtained from  
the first law of Thermodynamics as ${\cal P} = -\sum_\ell \Phi_\ell$. In particular,  
${\cal P} = 0$ when the $J_\ell$'s are equal.  

From Eq.~(\ref{eqg}), one obtains  
the ratio $w_{\ell i}(\eta)/w_{\ell i}(\eta^i)$ for each thermal reservoir $\ell$, consistent with the local detailed balance:
\begin{equation}
    \frac{w_{\ell i}(\eta)}{w_{\ell i}(\eta^i)} 
    = e^{-\beta_\ell [E_\ell(\eta^i) - E_\ell(\eta) - \theta_\ell F_\ell]}.
\label{23}
\end{equation}

By inserting Eq.~(\ref{23}) into Eq.~(\ref{eq:ep3})  
and taking into account that thermal reservoirs are mutually exclusive,  
the entropy production can be expressed as $\sigma = \sum_\ell \sigma_\ell$, where each entropy flux component $\sigma_\ell$ is given by
\beq
    \sigma_\ell = \sum_\eta p^{\rm st}(\eta) \sum_i
    w_{\ell i}(\eta) \ln \frac{w_{\ell i}(\eta)}{w_{\ell i}(\eta^i)}.
    \label{53_2}
\eeq
It is straightforward to see that the entropy flux component $\sigma_\ell$ is related to the exchanged heat $\Phi_\ell$ by a Clausius-like relation,  
$\sigma_\ell = -\beta_\ell \Phi_\ell$, where $\Phi_\ell$ is given by Eq.~(\ref{4}).

{\it Generic and Majority Vote Models----}  
We now turn to the relationship between this class of heat engines and opinion models. Although the thermodynamics of opinion systems is rather unconventional, their key features and phase transitions, above all the behavior of the entropy production
have been set up recently \cite{de1992isotropic,crochik2005entropy,encinas2019majority,noa2019entropy}. Moreover, it offers a simple way to relate the different temperatures governing the dynamics by means of a common parameter (probability rate).  

The generic vote model has a transition rate assuming the form  
$w_i(\sigma) = [1 - \sigma_i g(\ell)] / 2$, where $g(\ell)$  
depends on the neighborhood $\ell$ and is  
an odd function, constrained between $0$  
and $1$. Its simplest case is the  
majority vote (MV) model \cite{de1992isotropic},  
in which $g(\ell) = \mathrm{sgn}(\ell)$ for all $\ell$. Transition  
rates for the MV are then given by  
\begin{equation}
  w_i(\eta) = \frac{1}{2}\left\{1 - (1 - 2f) \eta_i \mathrm{sgn}(\ell)\right\},
\label{eq2}
\end{equation}  
meaning  
the spin $\eta_i$ tends to align itself with the local majority of its neighborhood with probability $1 - f$, and with complementary probability $f$, the majority rule is not followed. Eq.~(\ref{eq2}) is valid for all values of $f$ in the interval $0 \leq f \leq 1/2$.  

From Eq.~(\ref{eq2}), the ratio between $w_i(\eta)$  
and its reverse $w_i(\eta^i)$ is given by  
\begin{equation}
\ln \frac{w_i(\eta)}{w_i(\eta^i)} = - \eta_i \, \mathrm{sgn}(\ell) \ln \left(\frac{1 - f}{f}\right),
\label{iner}
\end{equation}

To obtain the relationship between $\beta_\ell$'s and $f$,
we take the energy difference due to the spin flip, given by $\Delta E=2 \eta_i |\ell|\mathrm{sgn}(\ell)$ for the Ising model by setting  $H_\ell=0$, such latter 
assumption because   $\mathrm{sgn}(\ell)=-\mathrm{sgn}(-\ell)$ for  any $\ell$. 
By choosing for simplicity
 $\theta_\ell=\eta_i {\rm sgn}(\ell)$ and
 taking the logarithm of Eq. (\ref{23}), it follows that \textbf{}
\begin{equation}
    \ln\frac{w_{\ell i}(\eta)}{w_{\ell i}(\eta^i)}=-\beta_\ell \eta_i \mathrm{sgn} (\ell)(2J_\ell|\ell|+ F_\ell).
\end{equation}
Since the transition rates associated with each {thermal reservoir}
are mutually exclusive, a direct comparison with Eq. (\ref{iner}) for a given $\ell$ provides the evaluation of each $\beta_\ell$ given by
\begin{equation}
    \beta_{\ell}=\frac{1}{(2J_\ell|\ell|+F_\ell)}\ln(\frac{1-f}{f}),
    \label{temp}
\end{equation}
where one has $\beta_2J_2=2\beta_4J_4=3\beta_6J_6...=k\beta_kJ_k/2$ for $F_\ell=0$
and $\beta_2(4J_2+F_2)=\beta_4(8J_4+F_4)=\beta_6(12J_6+F_6)...$ for $F_\ell\neq 0$. Note that the right-hand side of Eq.~(\ref{temp}) establishes a relationship among all the temperatures involved and will be taken into account in the subsequent analysis.

To link the present description to a generic voter model, it is  required the knowledge of $g(\ell)$
for each $\ell$. By taking for simplicity  $k=4$ and $F_\ell=0$, 
$g(\ell)$ can be expressed as  $g(0)=0,|g(2)|=q$
and $|g(4)|=p$. The relationship between  $\beta_2J_2(\beta_4J_4)$ with 
$q(p)$ can be obtained
as before. In particular, the condition  $\beta_2J_2=n\beta_4J_4$ between temperatures for $n>2$ implies that $p$
and $q$ are related via expression
\begin{equation}
p=\frac{\left(\frac{1+q}{1-q}\right)^{2/n}-1}{\left(\frac{1+q}{1-q}\right)^{2/n}+1}.
\end{equation}

In all cases, from Eq.~(\ref{4}) one has $\Phi_\ell=\sum_i(2J_\ell|\ell|+F_\ell)\langle\eta_i\mathrm{sgn}(\ell)w_{\ell i}(\eta)\rangle$ and hence  the  expression for ${\cal P}$ becomes
\begin{eqnarray}
{\cal P}&=&
  -\sum_\ell\sum_i(2J_\ell|\ell|+F_\ell)\langle\eta_i\mathrm{sgn}(\ell)w_{\ell i}(\eta)\rangle.
\label{p5}
\end{eqnarray}
A heat engine, providing partial conversion of energy from the  hot thermal reservoir(s) $\langle {\dot Q}_2\rangle=\sum_\ell \Phi_\ell H[\Phi_\ell]$ ($H[x]$ being the Heavside function) into power ${\cal P}<0$  has  associate  efficiency $\eta$  given by $\eta=-{\cal P}/\langle {\dot Q}_2\rangle$.
\section{Main results}\label{main}

{\it Two thermal reservoirs---} The simplest heat engine corresponds to the $k=4$ case, meaning that each unit has four nearest neighbors and is placed in contact with two different thermal reservoirs. By rewriting  
the heat flux in Eq.~(\ref{4}) as $\Phi_\ell = J_\ell \phi_\ell + F_\ell \xi_\ell$, where  
$\phi_\ell = 2 \sum_i \langle |\ell| \mathrm{sgn}(\ell) \eta_i w_{\ell i}(\eta) \rangle$  
and $\xi_\ell = \sum_i \langle \mathrm{sgn}(\ell) \eta_i w_{\ell i}(\eta) \rangle$,  
and taking into account that ${\cal P} = 0$ when $J_2 = J_4$ and $F_\ell = 0$, the first law  
of thermodynamics states that $\phi_2 = -\phi_4$, where $\phi_4 > 0$.  
The power is then written as ${\cal P} = (J_2 - J_4) \phi_4 - F_2 \xi_2 - F_4 \xi_4$.  
In order to investigate two different kinds of worksources, we split the analysis into two different cases: $J_2 \neq J_4$ and $F_\ell = 0$,  
and $J_2 = J_4 = J$ for $F_\ell \neq 0$. We pause to make  
a few comments about the former case. First, the efficiency $\eta$ acquires  
the simple form $\eta = 1 - J_2 / J_4$, which is lower than  
$\eta_c = 1 - \beta_4 / \beta_2$, also expressed as  
$\eta_c = 1 - J_2 / (n J_4)$ (because $\beta_2 J_2 = n \beta_4 J_4$). They are related via the relation $\eta = 1 - n + n \eta_c$,  
consistent with $\eta < \eta_c$ for  
all values of $J_2$ and $J_4$. Second, in contrast with ${\cal P}$ and $\langle Q_\nu \rangle$'s, $\sigma$ is independent of $J_2$ and  
$J_4$ and is given by  
$\sigma = -\beta_2 \Phi_2 - \beta_4 \Phi_4 = (\beta_2 - \beta_4) \phi_4$.  
Third, they are not independent of each other, but rather related through the identity  
$-\beta_2 {\cal P} \eta_c = J_4 \eta \sigma$. 
Due to the fact that $\sigma$ does not depend on the $J_\ell$'s,  
they can be adjusted to improve power and efficiency (although $\eta < \eta_c$) without sacrificing dissipation. Fourth and last, since $\phi_4 > 0$, the heat engine is characterized by $J_2 < J_4$. The roots of the power occur at $J_2 = J_4$ and $\phi_4 = 0$, the latter coinciding with the minimum of $\sigma$, in conformity with Ref.~\cite{forao2025characterization}.  
The case of $F_\ell \neq 0$ (exemplified here for $F_2 = F_4 = F \neq 0$ and $J_2 = J_4$) is less direct. From Eq.~(\ref{p5}),  
one has  
${\cal P} = -F_2 (\xi_2 + \xi_4)$ with efficiency $\eta = F(\xi_2 + \xi_4) / \Phi_4$, also bounded by the Carnot efficiency $\eta_c = 2 / (4 + F)$.  

\begin{figure}
\hspace{-0.4cm}
 \includegraphics[scale=0.32]{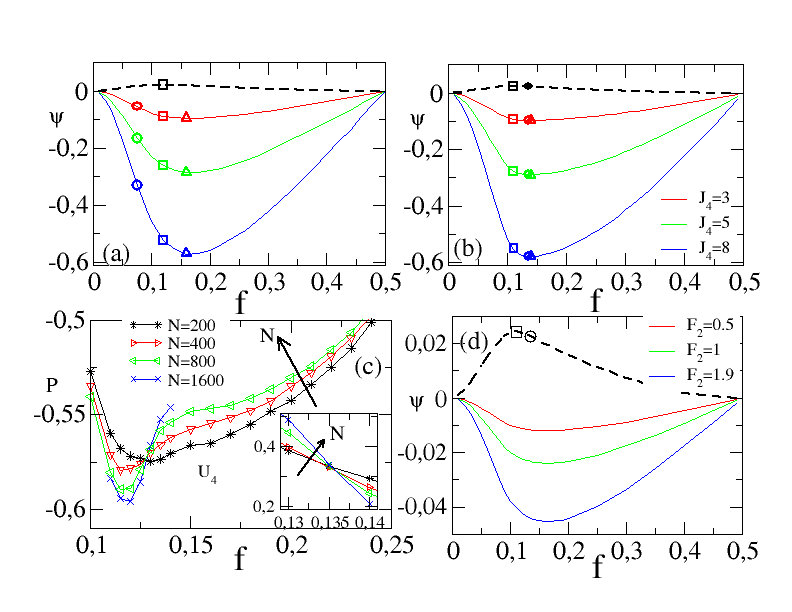} 
\caption{ Main results for the two thermal baths ($k=4$). Panels $(a)$ and $(b)$ depict for $N=100$ the power $\Psi={\cal P}$ (continuous lines) and the entropy production $\Psi=\sigma$ (dashed lines) for the MV
for square and RR  topologies, respectively, for $\beta_2J_2=2\beta_4J_4$ and different sets of $J_4$'s and $J_2=2$. Circles, squares and triangles denote $f_{c},f_{M\sigma}$ and 
$f_{mP}$, respectively.
Bottom left  panel shows the finite size scaling of power ${\cal P}$ for
RR topologies. Arrow indicates the increase of $N$. Inset: The location of critical point $f_c$ for $RR$ via crossing between curves of $U_4$ for different system sizes $N$.
The right (bottom) panel shows
some results ($\Psi={\cal P}$ and $\sigma$) for $J_2=J_4$ and distinct driving strengths $F_2$'s.
 In all cases, the efficiency $\eta$
is constant and given by $\eta=1-J_2/J_4$ (a-c) and $\eta=F(\xi_2+\xi_4)/\Phi_4$ (d).}
\label{fig1}
\end{figure}

Fig.~\ref{fig1} summarizes the above results for the  
$\beta_2 (4 J_2 + F_2) = \beta_4 (8 J_4 + F_4)$ case (MV model)  
for square and random-regular (RR) topologies, both with $k = 4$. For convenience,  
all analyses are depicted in terms of the parameter $f$ (temperatures are promptly evaluated from $f$ via Eq.~(\ref{temp})). In all cases, systems  
present continuous phase transitions as $f$ is raised.  
They are characterized via the standard finite-size scaling analysis, through the crossing between curves of the reduced cumulant $U_4 = 1 - \langle m^4 \rangle / (3 \langle m^2 \rangle^2)$ \cite{landau2021guide}, where $\langle m^{\bar n} \rangle$ denotes the ${\bar n}$-th order parameter,  
$\langle m^{\bar n} \rangle = \langle (\sum_{i=1}^N \eta_i / N)^{\bar n} \rangle$ ($N = L^2$ and $L^3$  
for square and cubic lattices, respectively). In particular,  
$\langle |m| \rangle$ and the entropy production derivative $\sigma^* = d\sigma / df$  
scale with  
$\langle |m| \rangle \sim L^{-\beta/\nu}$ and $\sigma^* \sim L^{\alpha/\nu}$ at the criticality $f = f_c$,  
respectively. Although $\beta/\nu = 1/8$ and $\beta/\nu = 1/2$ for square lattices and RR topologies, respectively \cite{encinas2018fundamental,encinas2019majority},  
$\alpha/\nu = 0$ in both cases, consistent with a logarithmic divergence \cite{noa2019entropy}.  
The value $f_{mP}$ at which $-{\cal P}$ is maximum  
is directly related to  
the maximum of $\phi_4$ and is different from $f_{M\sigma}$ at which $\sigma$  
is maximum.  
All such points lie  
in the interval $f_c < f_{M\sigma} < f_{mP}$  
for square-lattice ($f_c = 0.075(1)$ \cite{de1992isotropic}) topologies and change mildly as the  
system size is increased (not shown). On the other hand, the behavior for RR topologies is more sensitive to the system size (see e.g., Fig.~\ref{fig1}a–c), in which $f_{mP} \neq f_{M\sigma}$ and approaches $f_c$ (insets) as $N$ increases. Similar results are exemplified for  
different sets of $F_2 = F_4 \neq 0$ (Fig.~\ref{fig1}d).

\begin{figure}
\hspace*{-0.4cm}     
\includegraphics[width=\columnwidth]{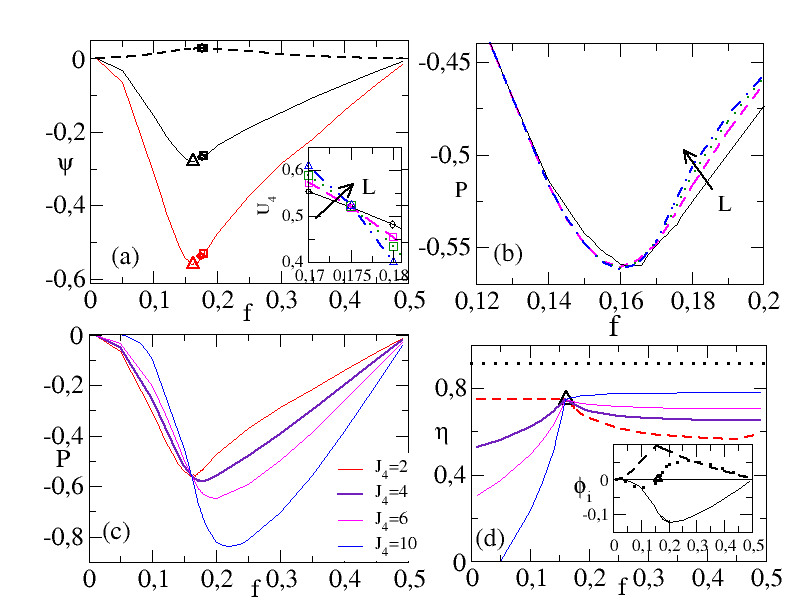} 
\caption{The depiction of power and efficiency for the  MV for a cubic ($k=6$) lattice for $\beta_2J_2=2\beta_4J_4=3\beta_6J_6$.
Panel $(a)$ shows, for $L=16~(N=16^3$ sites) , the behavior of ${\cal P}$ (continuous lines)
and $\sigma$ (dotted) for distinct $f$'s for $J_2=J_4=2$ and $J_6=5(8)$
for middle~(bottom) curves. Triangles, squares
and circles denote ${\cal P}_{mP}, {\cal}_{M\sigma}$ and ${\cal P}_c$, respectively.
Top inset: The location of $f_c$ via crossing between $U_4$ curves for different
system sizes.  Arrow indicates the increase of $L$.
In $(b)$ and $(c)$, the 
influence of distinct system sizes $L$'s and  intermediate $J_4's$ ($J_2=2$ and $J_6=8$), respectively, for distinct $f$'s.  The corresponding efficiency curves of panel c are reported
in d. Dotted line corresponds to the ideal efficiency $\eta_c$.  Bottom inset: The fluxes $\phi_\ell$'s versus $f$. Triangle denote the value
$f_0$ in which $\phi_4$ vanishes.}
\label{fig2}
\end{figure}

\begin{figure}
\hspace*{-0.4cm}     
\includegraphics[width=\columnwidth]{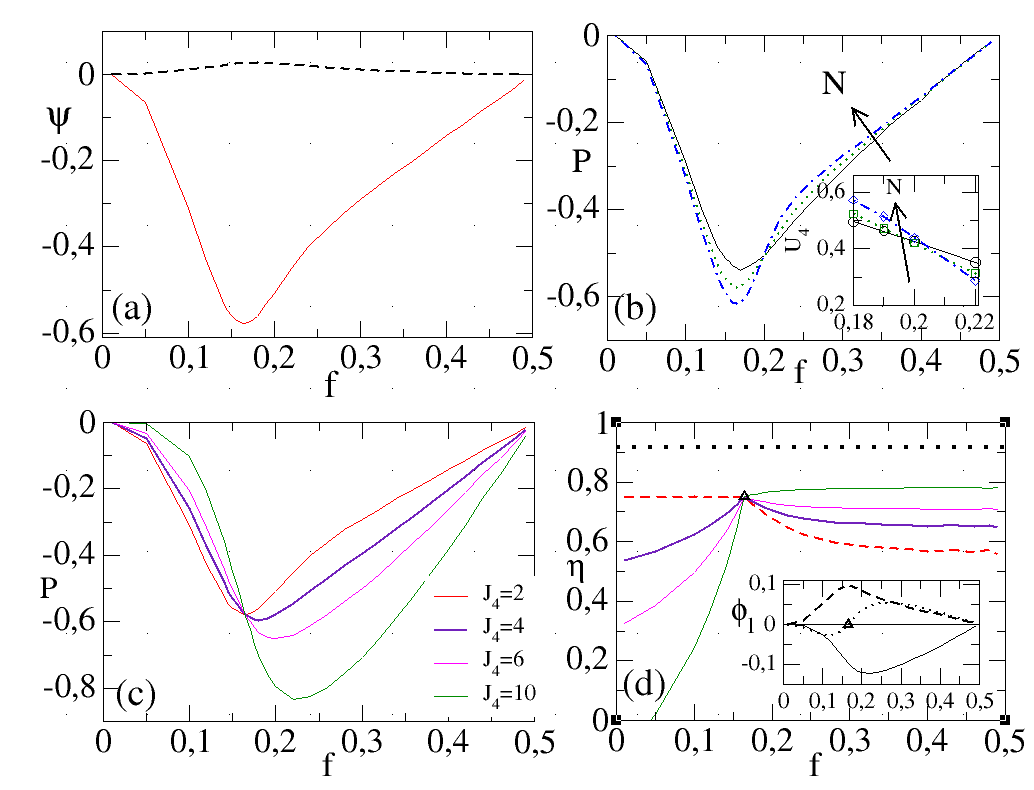} 
\caption{The same as Fig.~\ref{fig2} but for RR topologies as $k=6$.}
\label{fig3}
\end{figure}

{\it Three thermal reservoirs---} The $k=6$ case is more revealing  because
the system is placed in now contact with three different thermal reservoirs. 
As exemplified in Ref.~\cite{tome2022stochastic,felipe1}, the thermodynamic analysis of the MV shows that  $\phi_2<0$
and $\phi_{6}>0$ for all values of $f$, whereas the  intermediate heat flux ($\phi_4$ in such case) changes its sign at a intermediate $f_0$ (inset). By curbing ourselves in the case $F_\ell=0$, together the fact that $\phi_4=-(\phi_2+\phi_6)$, 
the expressions for thermodynamic quantities can be written down in the
following  forms, given by ${\cal P}=(J_4-J_2)\phi_2+(J_4-J_6)\phi_6, \langle Q_2\rangle=J_4\phi_4H[\phi_4]+J_6\phi_6$ (likewise $\langle Q_1\rangle=J_4\phi_4H[-\phi_4]+J_2\phi_2$), $\sigma=(\beta_4-\beta_2)\phi_2+(\beta_4-\beta_6)\phi_6$.
The efficiency  $\eta$ then reads
\begin{equation}
\eta=\frac{(J_2-J_4)\phi_2+(J_6-J_4)\phi_6}{J_4\phi_4H[\phi_4]+J_6\phi_6},
\label{eff}
\end{equation}
lower than $\eta\le \eta_c=1-J_2/(3J_6)$ (for the MV), for all
values of $J_\ell$'s (as should be). We see
that ${\cal P}$ presents three different roots,  
 at $J_2=J_4=J_6$ (absence of a heat-engine behavior), $\phi_2=\phi_6=0$ 
and $(J_4-J_2)/(J_4-J_6)=-\phi_6/\phi_2$, the second case corresponds to the minimum of dissipation for the MV  ($f=0$ or $f=1/2$). 
We also exemplify the three thermal baths case for 
for $\beta_2J_2=2\beta_4J_4=3\beta_6J_6$ (MV case),
as summarized in Figs.~\ref{fig2} and \ref{fig3}. As the two thermal baths previous case, the system also yields a continuous phase transition, 
presenting different critical exponents for regular (cubic) lattices
$\beta/\nu=5/16$ ($d=3$) \cite{salinas2001introduction} and $\beta/\nu=1/2$ for RR (like $k=4$) \cite{encinas2019majority}. 
Unlike the two thermal baths case, in which ${\cal P}$ curves solely differ by a factor $J_4-J_2$ with fixed $\eta$, the presence of an intermediate thermal reservoir can be conveniently chosen for improving the power and efficiency for both lattice topologies. Also, maximum powers $-{\cal P}_{mP}$'s also deviate for larger values $f$ (the system constrained in the disordered phase as $N\rightarrow \infty$) as $J_4$
increases. The behavior of $\eta$ is also different as a  consequence of $\phi_4$ changing its sign at $f_0$ ($f_0\approx 0.162$ for both cubic and RR, respectively), implying that
$\eta$ solely depends on $\phi_2$ and $\phi_6$  and given by \[
\eta=1-\frac{J_4}{J_6}+\frac{J_2-J_4}{J_6}\frac{\phi_2}{\phi_6}.\]
Note that $\eta=1-J_4/J_6$ for $J_2=J_4$, as depicted in panel
Figs.~\ref{fig2}d and \ref{fig3}d (dashed lines). By choosing suited values of $J_4$, the existence of an intermediate
thermal reservoir not only can improve the power, but also the efficiency, as depicted in Figs.\ref{fig2}d and \ref{fig3}d.


{\it Pair mean-field description for the power and dissipation---}
In order to provide additional insights about such class of heat engines beyond  numerical simulations, we investigate the system behavior via the mean-field
approach for $k=4$
by taking the correlation of two sites, also known
as pair mean-field approximation \cite{tome2015stochastic,tome2022stochastic}. At this level of approximation, all quantities can be expressed in terms of 
$m=\langle \eta_i\rangle$ and the two-site correlation.
and
$r=\langle\eta_0\eta_i\rangle$.
To see this, let
$P(\eta_1,\eta_2,\eta_3,\eta_4|\eta_0)$, $\eta_0$ and $\eta_i$ being
the spin of the central site and   its $k=4$ nearest neighbors,
respectively. At this level of approximation, the conditional probability is writing  as
$P(\eta_1,\eta_2,\eta_3,\eta_4|\eta_0)
= \prod_i P(\eta_i|\eta_0)$,
where
$P(\eta_i|\eta_0)=P(\eta_i,\eta_0)/P(\eta_0)$. They are related with $m$ and $r$ through  relations
 as $P(\eta_0) = (1+m\eta_0)/2$
and 
$P(\eta_i,\eta_0) = [1+m(\eta_0+\eta_i)
+ r\eta_0\eta_i]/4$, respectively.  In order to find
the steady solutions for thermodynamic quantities, we take
the time evolution of $m$ and $r$, expressed  in terms
of  correlations of type $\langle \eta_0\eta_1\ldots\eta_n\rangle$
and  $\langle \eta_1\ldots\eta_n\rangle$.
From Ref.~\cite{tome2022stochastic}, we see they can be conveniently written in the following forms $\langle \eta_0\eta_1\ldots\eta_n\rangle=(A_n-B_n)/2$
 and $\langle \eta_1\ldots\eta_n\rangle =(A_n+B_n)/2$, respectively, where
 \begin{equation}
A_n=\frac{(m+r)^n}{(1+m)^{n-1}}\qquad {\rm and}\qquad B_n=\frac{(m-r)^n}{(1-m)^{n-1}}.
\end{equation}
We arrive at the following equations for $m$ and $r$:
\beq
\frac{dm}{dt} = -m -4a m + 2b (A_3+B_3),
\label{47}
\eeq
and 
\beq
\frac{dr}{dt} = -2 r +2a +3(a+b)(A_2+B_2) + b (A_4+B_4),
\label{47a}
\eeq
respectively. Parameters $a$ and $b$ are simpler
for $\beta_2J_2=2\beta_4J_4$ (MV for $F_\ell=0$) and given by $b=-3a$, where $b=-(1-2f)/8$. For the generic case $\beta_2J_2=n\beta_4J_4$, one has
$a=(p+2q)/8$ and $b=(p-2q)/8$, where $p$ and $q$ were described previously.
As stated previously, quantities can be solely
expressed in terms of $\phi_4$ for $k=4$, 
${\cal P}=(J_2-J_4)\phi_4$ and $\sigma=(\beta_2-\beta_4)\phi_4$, where flux $\phi_4$ reads
\begin{equation}
\phi_4=A_3-B_3-a(A_4+B_4)-3(a+b)(A_2+B_2)-2b.
\end{equation}
The steady state solution characterizing  the ordered phase is found by solving numerically 
Eqs. (\ref{47}) and (\ref{47a}). On the other hand, they
become simpler in the disordered phase, in which  $m=0$ and
  $r=a+3(a+b)r^2+br^4$ and  the flux $\phi_4$ is given by $\phi_4=2 r^3-6 r^2 (a+b)-2 a r^4-2 b$
  or equivalently $\phi_4=\left[8 r^3-p \left(r^4+6 r^2+1\right)-2 q \left(r^4-1\right)\right]/4$.

\begin{figure}   
\includegraphics[width=\columnwidth]{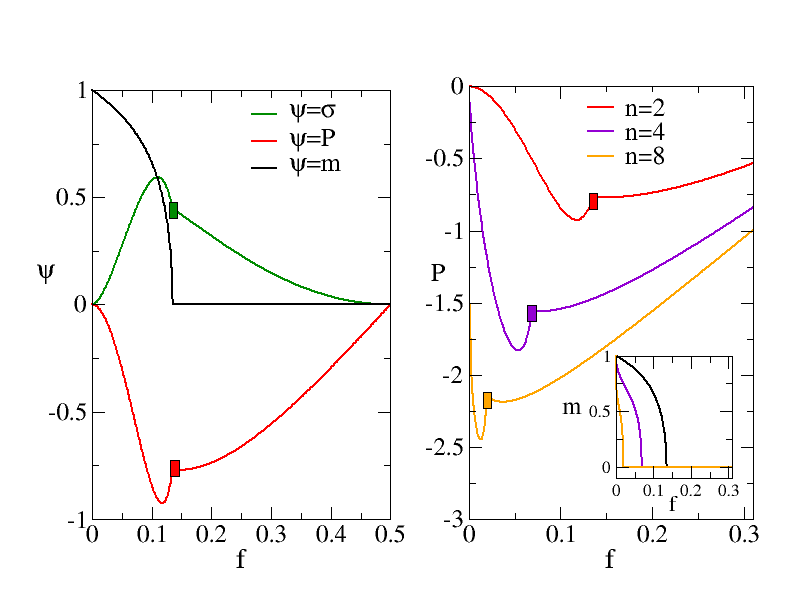} 
\caption{Left: For $\beta_2J_2=2\beta_4J_4$ (MV) and $F_\ell's=0$, the  depiction of  $\Psi=m,{\cal P}$ and  $\Psi=\sigma$ for the pair-approximation for $k=4$. 
Right: The comparison between ${\cal P}$'s for different $n$'s. Quantities have been expressed in terms of $f=(1-q)/2$.  The behavior of thermodynamic quantities  is marked by a kink (symbol $\blacksquare$)
at the phase transition, 
the disordered phase being at the right. Inset: The corresponding behavior for $m$'s. Parameters: $J_2=2$ and $J_4=8$.}
\label{fig3RR}
\end{figure}

  Fig.~\ref{fig3RR} summarizes above discoveries 
for $m,{\cal P}$ and $\sigma$ for $n=2$ as well as
a comparison of ${\cal P}$ for distinct $n's$.
As can be seen,  it reproduces previous similar trademarks of a heat-engine in both
ordered and disordered phases, in which both maximum power and dissipation yields at the ordered phase, consistent with the collective behavior
 improving the system performance. Although close to each other,
$f_{mP}$  does not coincide with $f_{M\sigma}$. Contrasting with previous cases ${\cal P}$ and $\sigma$ are marked by a kink
 at the phase transition.  Also,
both ${\cal P}$ and its maximum values $-{\cal P}_{mP}$
increase as $n$ (larger difference of temperatures) goes up. 
However, the ordered phase is substantially shortened in this case. 

\section{Conclusions}\label{cond}
We introduced an alternative kind of nonequilibrium thermal engines operating
collectively, based on
the idea of a well defined thermal reservoir due to the spin neighborhood.   Unlike other kinds of collective heat engines \cite{powerful}, the system  exhibits heat engine behavior in both ordered and disordered phases. We derived a general relation between power, efficiency and dissipation
for the two thermal reservoir case. The three thermal baths exhibit richer features in which the intermediate thermal reservoirs can be adjusted ensuring the improving of power and efficiency. Contrasting
with other examples of collective systems \cite{mamede23}, our results indicate
that topology
of interactions does not play a relevant role in the system performance.
The connection with different  kinds of voter models was also considered
and exemplified, highlighting different routes for improving the power and efficiency.

\section{Acknowledgements}
 We acknowledge Gustavo For\~ao for the reading of the manuscript 
 and insightful suggestions. The financial support from Brazilian agencies CNPq and FAPESP under grants 2023/17704-2, 2024/08157-0, 2024/03763-0, 2022/15453-0
 is also acknowledged.

\bibliography{refs,refs2}
\end{document}